\documentclass[a4paper,twocolumn,11pt,unpublished]{quantumarticle}
\pdfoutput=1
\usepackage[utf8]{inputenc}
\usepackage[english]{babel}
\usepackage[T1]{fontenc}
\usepackage{amsmath}
\usepackage{bbold}
\usepackage{amssymb}
\usepackage{hyperref}
\usepackage{bm}
\usepackage{dsfont}
\usepackage{amsthm}
\usepackage{subcaption}
\usepackage{tikz}
\usepackage{lipsum}

\newcommand{\rang}{\rangle}
\newcommand{\lang}{\langle}

\newtheorem{definition}{Definition}
\DeclareMathOperator*{\argmin}{argmin}

\newtheorem{theorem}{Theorem}

\begin{document}

\title{Quantum $k$ nearest neighbors algorithm}

\title{Quantum $k$ nearest neighbors algorithm}
\author{Afrad Basheer}
\email{Afrad.M.Basheer@student.uts.edu.au}
\affiliation{Centre for Quantum Software and Information, University of Technology, Sydney, Australia}
\affiliation{Chennai Mathematical Institute,
H1 SIPCOT IT Park, Kelambakkam, Tamil Nadu 603103, India}
\author{A. Afham}
\email{Afham@student.uts.edu.au}
\affiliation{Centre for Quantum Software and Information, University of Technology, Sydney, Australia}
\affiliation{Department of Physical Sciences, Indian Institute of Science Education \& Research (IISER) Mohali, Sector 81 SAS Nagar, Manauli PO 140306 Punjab India.}
\author{Sandeep K.~Goyal}
\email{skgoyal@iisermohali.ac.in}
\affiliation{Department of Physical Sciences, Indian Institute of Science Education \& Research (IISER) Mohali, Sector 81 SAS Nagar, Manauli PO 140306 Punjab India.}

\maketitle

\begin{abstract}
  One of the simplest and most effective classical machine learning algorithms is the $k$ nearest neighbors algorithm ($k$NN) which classifies an unknown test state by finding the $k$ nearest neighbors from a set of $M$ train states. Here we present a quantum analog of classical $k$NN -- quantum $k$NN (Q$k$NN) -- based on fidelity as the similarity measure. We show that the Q$k$NN algorithm can be reduced to an instance of the quantum $k$ maxima algorithm; hence the query complexity of Q$k$NN is $O(\sqrt{kM})$. The non-trivial task in this reduction is to encode the fidelity information between the test state and all the train states as amplitudes of a quantum state. Converting this amplitude encoded information to a digital format enables us to compare them efficiently, thus completing the reduction. Unlike classical $k$NN and existing quantum $k$NN algorithms, the proposed algorithm can be directly used on quantum data, thereby bypassing expensive processes such as quantum state tomography. As an example, we show the applicability of this algorithm in entanglement classification and quantum state discrimination.
\end{abstract}

\section{Introduction} \label{Sec:Introduction}

Quantum machine learning (QML)~\cite{Biamonte2017, Wittek2014, Schuld2014, Arunachalam2017} is a recent offspring of quantum computing and machine learning (ML). One of the proposed applications of QML is using quantum computers to speed up ML tasks~\cite{Harrow2009, Wiebe2012}. In a converse manner, ML has proven itself adept at problems in physics~\cite{Carrasquilla2017, Wang2017, Lu2018, Carleo2019}. We have also seen the realisation of quantum versions of several (classical) ML algorithms~\cite{Lloyd2014, Rebentrost2014, Lloyd2013}. Along the same vein, we propose a quantum version of the $k$ nearest neighbor algorithm~\cite{Cover1967} by reducing it to an instance of the quantum $k$ maxima finding algorithm~\cite{Durr2006, Miyamoto2019} and providing an explicit construction of the oracle required. The algorithm is capable of classifying states without their explicit classical description. We then present two problems of interest as applications of our algorithm, namely entanglement classification and quantum state discrimination.

$k$NN is a simple supervised ML algorithm used extensively for pattern recognition and classification. This algorithm rests on the assumption that two points close to each other are more likely to be of the same type. The algorithm is provided with a set of train states (vectors) whose class labels are known. The test state (vector) with the unknown label is compared with the train states, and the $k$ number of nearest neighbors of the train states are identified for the given test state. Finally, the label of the test state is determined upon majority voting.

The expensive step in the $k$NN algorithm is to determine the distance between the test state and all the train states. Each state (train or test) is represented by a vector of real (or complex) numbers. As the number of train states and the size of the state vectors increases, $k$NN becomes more expensive. To classify vectors of dimension $N$ by comparing it to a set of train vectors of cardinality $M$, we need to carry out $O(MN)$ operations.

    \begin{figure*}
        \center 
            \includegraphics[scale = 0.8]{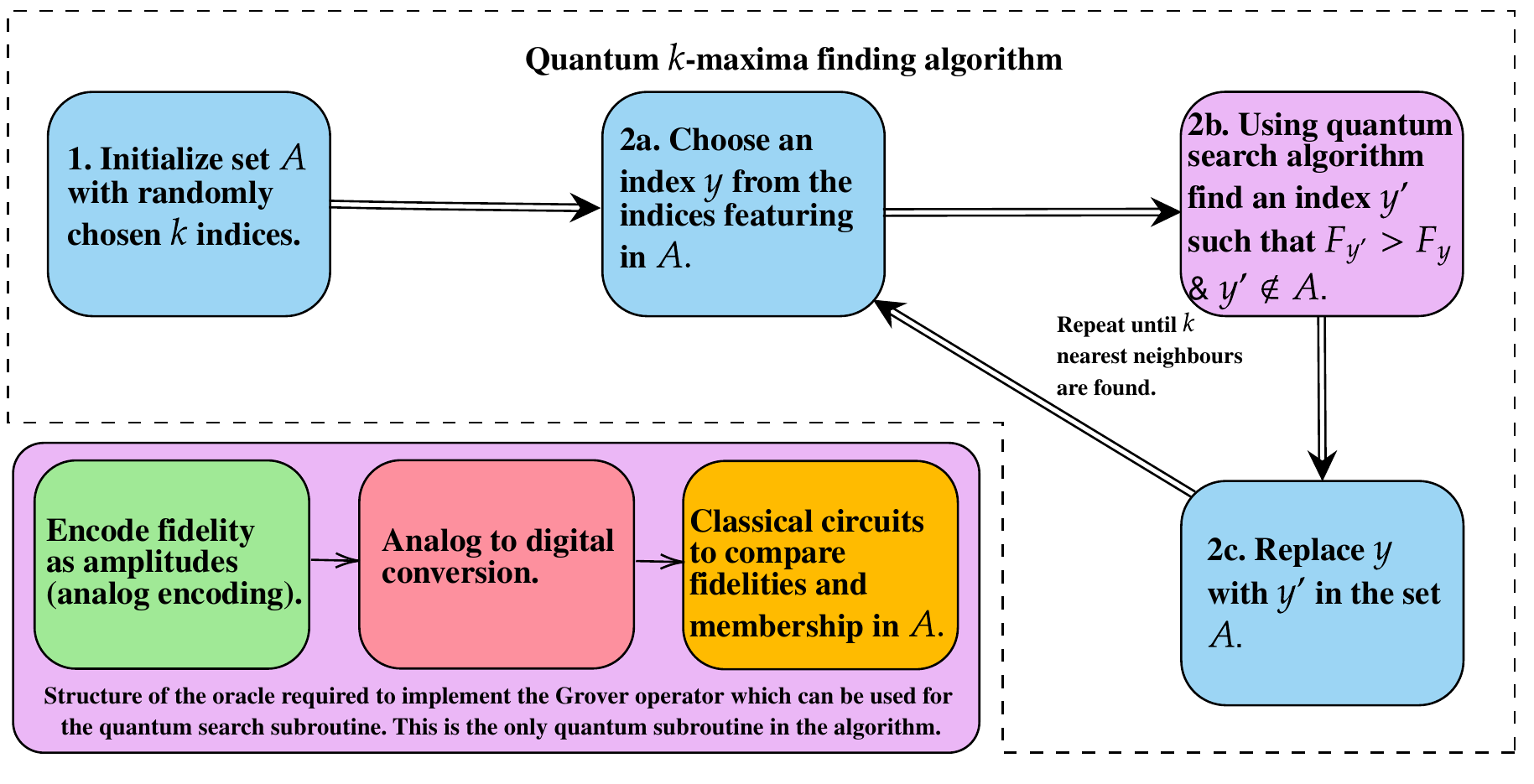}
            \caption{An overview of the  Q$k$NN algorithm. The idea is to use the quantum $k$ maxima finding algorithm on a table $F = [F_0,\ldots, F_{M-1}]$ to find the $k$ nearest neighbors of the test state. Here $F_j = |\lang \psi|\phi_j\rang|^2$ is the fidelity of the test state $|\psi\rang$ with the $j$th train state $|\phi_j\rang$. We start with a set $A$ of randomly chosen $k$ indices and then, using quantum search algorithm (Grover's search when the number of targets are unknown), we replace indices in $A$ with new indices (that are not in $A$) until the top $k$ neighbors are found. The crucial step here is to prepare an oracle capable of performing the required Grover operator in the quantum search subroutine. The oracle should be capable of making comparisons of the form $F_{y'} > F_y$ and ensure that $y' \notin A$. Such an oracle is constructed by first extracting the fidelity using Swap test and encoding it as amplitudes of a quantum state. Then, an analog to digital conversion of the amplitudes is carried out which results in the fidelity being encoded as digital (bit-string) states. Once this is done, we perform a series of classical operations to compare the fidelities and check membership in $A$.}\label{fig:flowchart}
    \end{figure*}

In this article, we propose a novel quantum $k$ nearest neighbor (Q$k$NN) algorithm, a quantum analog of the classical $k$NN algorithm. Utilising existing algorithms such as the Swap test~\cite{Buhrman2001}, $k$ minima finding algorithm~\cite{Durr2006}, quantum phase estimation, and the recent quantum analog-to-digital conversion algorithm~\cite{Mitarai2019}, we construct an algorithm capable of classifying states without the requirement of having their classical description (writing out the amplitudes in some basis). This is particularly useful in cases where the data to be classified is inherently quantum, and thereby we can bypass expensive processes such as tomography~\cite{Aaronson2007} to learn the description of the state in question. We also provide two applications of our algorithm, namely entanglement classification of pure states and quantum state discrimination. Moreover, since we use quantum $k$ maxima finding algorithm to find the $k$ nearest neighbors, the query complexity of our algorithm is $O(\sqrt{kM})$.

Previous work in quantum versions of $k$NN include \cite{Wiebe2015, Ruan2017, Chen2015, Dang2018, Schuld2014b}. Ruan et al.~\cite{Ruan2017} uses Hamming distance as the metric to estimate the $k$ nearest neighbors. Dang et al.~\cite{Dang2018} uses quantum $k$ nearest neighbors algorithm proposed in~\cite{Chen2015} for image classification. Of all the previous papers mentioned, the work by Chen et al.~\cite{Chen2015} seems to be the closest to ours with the same overall query complexity of $O(\sqrt{kM})$. 

However, there are a few features that distinguish our work from the above-mentioned works. Firstly, we envision our algorithm to be used directly on quantum data (though it can also be used on $\lq$classical data'), which allows us to classify states without having their explicit classical description. Instead, we require the circuits capable of preparing the test state. Also, we use fidelity and dot product as a measure of similarity, and we demonstrate problems where fidelity can be used to carry out classification.

In Section \ref{Sec:background}, we present the necessary background, and in Section \ref{Sec:QKNN}, we present our quantum $k$ nearest neighbors algorithm, which uses fidelity as a measure of closeness. Then, in Section \ref{sec:qknndot}, we present a variation of the algorithm that uses dot product instead of fidelity for classification. Both of these algorithms build upon generalisations of algorithms from~\cite{Mitarai2019}. Section \ref{sec:comp} comprises the query complexity of our algorithm which we show to be $O(\sqrt{kM})$. 
    We then present in Section \ref{sec:app} two problems where our  Q$k$NN algorithm can be utilised - the problem of entanglement classification and a problem analogous to quantum state discrimination. 
      
    Finally, we conclude in Section \ref{sec:conclusion} while also discussing the kind of problems our algorithm would be adept at solving.
    
\section{Background}\label{Sec:Background}
    \label{Sec:background}
    In this section, we provide the relevant background for the Quantum $k$NN algorithm. We begin with the introduction of the classical $k$NN algorithm
    \subsection{Classical $k$NN algorithm}
    
        Let $\{u_n\}$ be a collection of vectors of unknown labels, which we call \textit{test states}. The aim is to accurately assign these test states labels. For this, the $k$ nearest neighbors ($k$NN) algorithm requires a collection of vectors $\{v_m\}$ of the same dimension whose labels are known to us. We shall call these states \textit{train states}. $k$NN assigns labels to each test state by first computing the $k$ nearest neighbors of the test state. Then a majority voting is carried out among these $k$ nearest neighbors. Ties are resolved in different ways, such as assigning the label of the nearest training point or the label of a random training point among the $k$ nearest neighbors. Successful applications of $k$NN include~\cite{Liao2002, Zhang2003}. Being a simple algorithm, $k$NN also allows us to reason about the structure of the data we are working with. 
                
         Let the test states and the trains states are $r$-dimensional real or complex vectors. Any bona fide definition of a distance measure can be used for the purpose of $k$NN algorithm. Most common distance measures include Euclidean distance $\text{d}(\textbf{u},\textbf{v})=\sqrt{\sum_{i}^r|u_i - v_i|^2}$ and cosine similarity $\text{c}(\textbf{u},\textbf{v})=\frac{\lang \textbf{u}|\textbf{v}\rang}{\|\textbf{u}\| \cdot \|\textbf{v}\|}$ (which reduces to inner product for normalised states), where $\textbf{u}$ and $\textbf{v}$ are $r$-dimensional complex vectors.
     
         In quantum information, the fidelity function, though not a metric, $\text{F}(\cdot,\cdot)$ can be used to assign a notion of nearness between two quantum states beloning to the same space. For arbitrary quantum states $\rho, \sigma$ belonging to the same space,the fidelity between them is $\text{F}(\rho,\sigma)  = \left(\text{Tr}\left(\sqrt{\sqrt{\rho}\sigma \sqrt{\rho}}\right)\right)^2$. For pure quantum states $\rho = |u\rang \lang u |, \sigma = |v\rang \lang v |$, the fidelity function simplfies to $\text{F}(u,v) = |\lang u|v\rang|^2$.
         
        Though fidelity is not a metric, the metric Bures distance $\text{B}(u,v)~=~\sqrt{2 - 2\sqrt{\text{F}(u,v)}}$, defined over quantum states, is a monotonous function of the fidelity. Therefore, finding the \textit{k}-nearest neighbors of any quantum state with respect to the Bures distance is the same as finding the \textit{k} states with the largest fidelity to the chosen quantum state.
        
\begin{figure}
             \begin{center}
                 \includegraphics[scale=0.28]{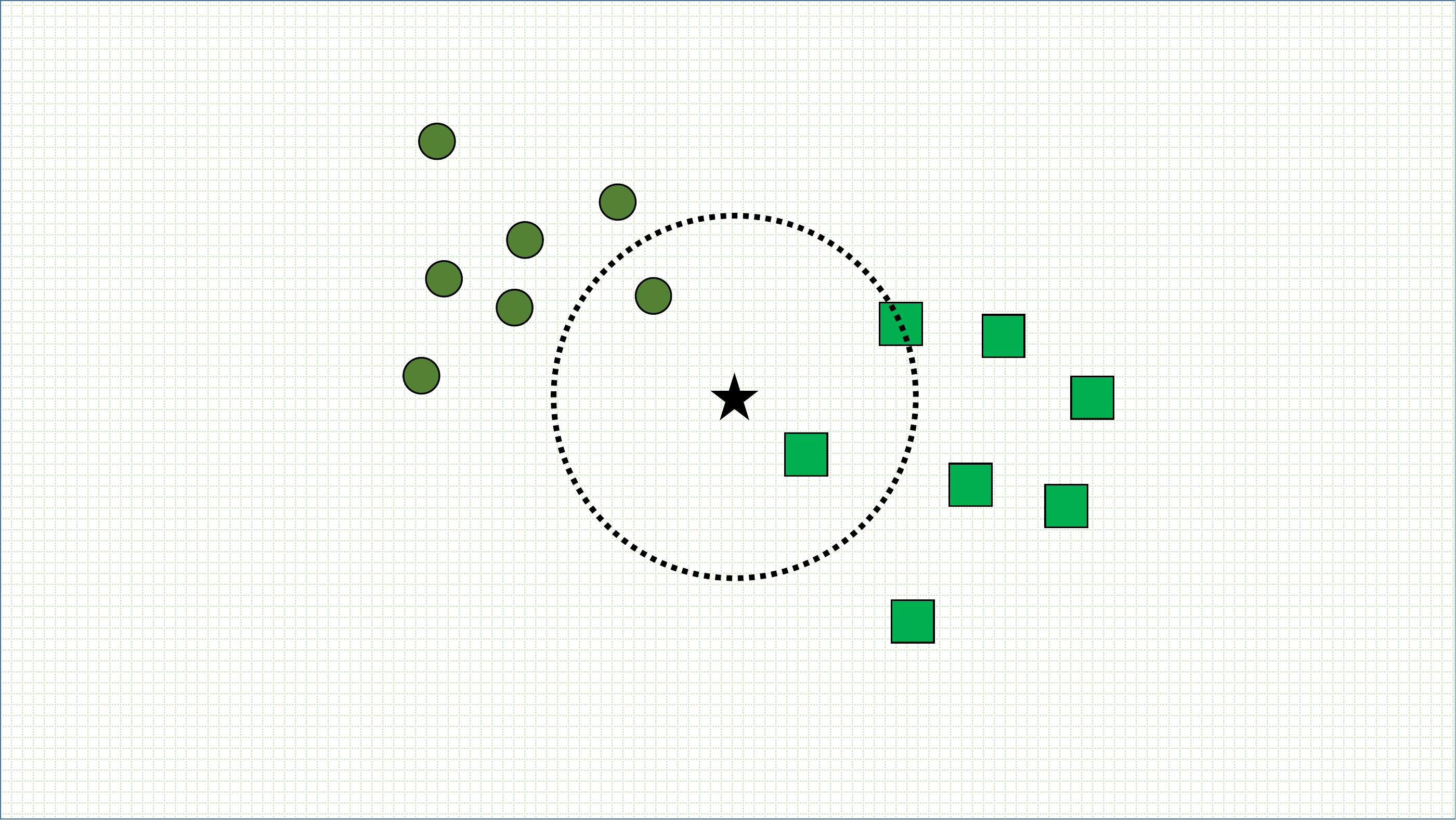}
                 \caption{Choosing a $k$ = 3 neighborhood. Here circle and square represents two different classes and star represents the unknown state whose label is to be determined. On choosing $k=3$, we classify it as a $\lq$square' point.}
                 \label{fig:knn}
             \end{center}
         \end{figure}

        The rationale behind $k$NN is that data points that are close together, with respect to some distance measure, must be similar. Formally, the $k$NN algorithm consists of the following steps (see Fig. \ref{fig:knn}):
        \begin{enumerate}
             \item For each test state (whose label is to be determined), compute its distance to the train states whose labels are known.
             \item Choose the $k$ number of neighbors which are nearest to the test state.
             \item Assign the label using majority voting. 
        \end{enumerate}
        
         Although the $k$NN algorithm is simple to understand and easy to implement, the algorithm has its drawbacks. As the number of train data points and the dimension of the state vectors grows, $k$NN can quickly turn intractable. Classification of an $N$ dimensional test state by comparing with $M$ train states requires $O(MN)$ multiplication operations. Furthermore, there is no general way of choosing $k$, and usually, hyperparameter tuning is done to choose the best possible $k$~\cite{Samworth2012}.

        \subsection{Quantum $k$ maxima finding algorithm}
        
        Durr and Hoyer describe, based on Grover search algorithm, an algorithm in \cite{Durr1996} which can be used to find the minimum of an unsorted list of size $M$ with complexity $O(\sqrt M)$. Through a simple modification, one can use the same algorithm to find the maximum instead of the minimum.  A generalisation of the algorithm can be found in~\cite{Durr2006}, which can be used to find the $k$ smallest elements in a table $T = [T_0,\ldots, T_{M-1}]$ of $M$ elements in time $O(\sqrt{kM})$. A simple explanation of the algorithm can be found in~\cite{Miyamoto2019}.

        The general idea of the algorithm is to start with randomly chosen $k$ indices and use the quantum search algorithm~\cite{Boyer2005} to find and replace the chosen indices with ones that have a higher table value. This process is repeated until we end up with the $k$ highest values in the table.
        The algorithm is as follows: 
        \begin{enumerate} \label{algo2}
            \item Initialise a set $A = \{i_1, \ldots, i_k\}$ with randomly chosen $k$ indices from the list of $M$ indices.

            \item Repeat the following forever: \label{main_loop}
            
            \begin{enumerate}
                \item Select threshold index $y$ from $A$ randomly. \label{kmax_step1}
                \item \label{QkMax} Using quantum search algorithm proposed in~\cite{Boyer2005}, find index $y'$ which is not present in $A$, such that $T_{y'} > T_{y}$. This can be seen as using quantum search on the Boolean function
                \begin{equation}
                    f_{y,A}(j)  = 
                    \begin{cases}
                        1 &: T_j > T_y \text{ and } j \notin A,\\
                        0 &: \text{otherwise}.                        
                    \end{cases}
                \end{equation}
                \label{sotp_crit}
                \item Replace $y$ with $y'$ in the set $A$. 
            \end{enumerate}
        \end{enumerate}
        The step \ref{QkMax} is the only quantum subroutine of the algorithm. Essentially, we use quantum search to find an index not present in $A$ and has its table value greater than the table value of the threshold index. The quantum search algorithm is a slightly modified Grover's search algorithm, which is used in cases where the number of solutions to the boolean function is unknown. Note that this algorithm also uses the same Grover operator as in normal Grover's search. Also, instead of randomly sampling an index from $A$ in step \ref{kmax_step1}, one can choose the index \[y = \argmin_{i \in {A}} T_i.\]
        This method would ensure a stopping criterion for the algorithm, that is, repeat step \ref{main_loop} until step~\ref{sotp_crit} cannot be carried out. This is because if we cannot find an index, which is not in $A$ and has a higher table value than the index with the minimum table value in $A$, then we should have all the $k$ indices with the largest table values already present in $A$.
    
    \subsection{Swap Test}\label{app:SwapTest}
    
    \begin{figure}
            \begin{center}
                \includegraphics[page=3]{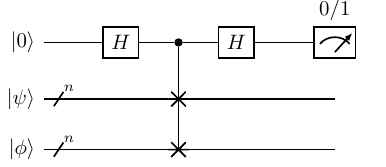}
                \caption{Circuit for Swap test.}
                \label{fig:swap}
            \end{center}
        \end{figure}

         The swap test~\cite{Buhrman2001} is a quantum algorithm that can be used to statistically estimate the fidelity $\text{F}(\psi,\phi) = |\langle\psi|\phi\rangle|^2$ between two arbitrary $n$ qubit pure states $|\psi\rangle$ and $|\phi\rangle$. The three register gate in Fig. \ref{fig:swap} is the \textit{controlled swap} (CSWAP) gate whose action is defined by
         
         \begin{align}
         \begin{split}    
             \text{CSWAP} |0\rang |\psi\rang |\phi\rang = |0\rang |\psi\rang |\phi\rang, \\ 
             \text{CSWAP} |1\rang |\psi\rang |\phi\rang = |1\rang |\phi\rang |\psi\rang.
        \end{split}     
         \end{align}
         
        To implement the swap test between states $|\psi\rang$ and $|\phi\rang$, we need three registers prepared in states $|0\rang$, $|\psi\rang$, and $|\phi\rang$, respectively. The initial combined state of the three registers is $|{0}\rang|{\psi}\rang|{\phi}\rang$. We then implement circuit in Fig. \ref{fig:swap}.

        At the end of the circuit, the measurement probabilities of the first register are
        \begin{align}
          \text{Pr}(0) & = \frac{1}{2} + \frac{1}{2}|\langle \psi|\phi\rangle|^2,\\
          \text{Pr}(1) & = \frac{1}{2} - \frac{1}{2}|\langle \psi|\phi\rangle|^2.
        \end{align}
        The quantity $\text{Pr}(0)- \text{Pr}(1)$ gives us the desired fidelity.
    \subsection{Quantum Analog to Digital Conversion (QADC) algorithm}
    Mitarai et al.~ describe a set of algorithms in~\cite{Mitarai2019} to carry out analog to digital conversions within a quantum circuit. We provide a description of what the algorithm does here and refer to Appendix (\ref{app:QADC}) for details for the sake of brevity.
    
    Let $\sum_{i=0}^{d-1} c_i |i\rang$ be an arbitrary quantum state. Let $\{r_0, \ldots, r_{d-1}\}$ be bitstrings that denote the best $b$-bit approximation of $\{|c_0|, \ldots, |c_{d-1}|\}$ respectively. An $m$-bit abs-QADC algorithm can transform the \textit{analog encoded} state $\sum_{i=0}^{d-1} c_i |i\rang$ to the \textit{digital encoded} state $\frac1{\sqrt{M}}\sum_{i=0}^{d-1} |i\rang |r_i\rang$. 
    
    Let $\{r_0, \ldots, r_{d-1}\}$ be bitstrings that denote the best \textit{b}-bit approximation of $\{\text{Re}(c_i), \ldots, \text{Re}(c_{d-1})\}$ respectively. An $m$-bit real-QADC algorithm transforms the \textit{analog encoded} state $\sum_{i=0}^{d-1} c_i |i\rang$ and to the \textit{digital encoded} state $\frac1{\sqrt{M}}\sum_{i=0}^{d-1} |i\rang |r_i\rang$. 

    In the coming sections, we show that variations to the circuits of these algorithms give rise to circuits capable of carrying out a quantum $k$NN algorithm.

\section{Quantum $k$ Nearest neighbors algorithm using fidelity}\label{Sec:QKNN}
    
    We now present the Quantum $k$ Nearest neighbors algorithm. Refer Fig. \ref{fig:flowchart} for an overview of the algorithm. Let $\mathcal{H}$ be the $n$-qubit Hilbert space of dimension $N = 2^{n}$ and let $|\psi\rang \in \mathcal H $ be the unknown \textit{test state} whose label is to be determined. Let  
    \begin{equation}
        \{|\phi_j\rangle : j \in \{0, \ldots, M-1\}\} \subset \mathcal H
    \end{equation}
    be a collection of $M$ \textit{train states} whose labels are known to us. For the sake of convenience, we assume $M = 2^m$ for some positive integer $m$. The idea is to find the $k$ nearest neighbors of $| \psi \rang$ from the train states and then through majority voting, assign $| \psi \rang$ a label. Let $F_j \equiv \text{F}(\psi, \phi_j) = |\lang \psi|\phi_j \rang|^2$ be the fidelity between the test state $|\psi\rang$ and the $j^{th}$ train state $|\phi_j\rang$ and define
    \begin{equation}
        F = [F_0, \ldots, F_{M-1}]
        \label{Ftable}
    \end{equation}
    to be a table of length $M$ containing the fidelities with the test state $|\psi\rangle$ and all the train states $\{| \phi_j \rang\}$. 
    
    \subsection{A summary of the algorithm}
    Note that the problem of finding the $k$ nearest neighbors of a test state $|\psi\rang$ can be reduced to an instance of $k$ maximum finding algorithm carried out on the table $F$ given in Eq.~\eqref{Ftable}. The only step that requires a quantum circuit in the quantum $k$ maximum finding algorithm is the quantum search subroutine given in step \ref{QkMax}. To achieve this, we should be able to prepare a circuit that carries out the oracle transformation 
    \begin{equation}\label{OracleOFid}
            \mathcal O_{y,A} |j\rang |0\rang = |j\rang |f_{y,A}(j) \rang.
        \end{equation}
    where $f_{y,A}$ is the Boolean function defined as 
        \begin{equation}\label{eq:fya}
            f_{y,A}(j) = 
            \begin{cases}
                    1 &: F_j>F_y \text{ and } j \notin A,\\
                    0 &: \text{otherwise}.
            \end{cases}
        \end{equation}
        That is, the fidelity $F_j$ must be greater than $F_y$ and $j$ should not feature in the threshold index set $A$. Q$k$NN algorithm using fidelity as the similarity measure is then:
        \begin{enumerate}
            \item
            Using $\mathcal O_{y,A}$ as the required oracle for the Grover operator in quantum search in step \ref{QkMax}, use $k$ maxima finding algorithm to find the $k$ indices $\{j_1, \dots ,j_k\}$ whose states $\{|\phi_{j_1}\rang, \dots ,|\phi_{j_k}\rang\}$  have the maximum fidelity with the test state. 
        
            \item
            Conduct a majority voting among the $k$ states and assign $|\psi\rang$ the label of the majority.     
        \end{enumerate}
    The non-trivial part in the above steps is the realisation of the oracle $\mathcal O_{y,A}$. We briefly discuss this oracle in the next subsection and provide an explicit construction of it after that.
    
    \subsection{On the oracle $O_{y,A}$}
    The construction of this oracle is based on the abs-QADC circuit from~\cite{Mitarai2019}. In that work, to compute the absolute values of the amplitudes of the state, we apply Swap test in superposition with the state and computational basis vectors. In the quantum $k$NN setting, we apply Swap test in superposition with the test state and the train states. Roughly speaking, the way to construct the oracle $\mathcal O_{y,A}$ is as follows:
    \begin{enumerate}
        \item \label{briefstep1} Construct an operator $\mathcal F$ capable of the transformation
            \begin{equation} \label{F_j_oper}
                \mathcal F |j\rang |0 \rang =  |j \rang |F_j\rang
            \end{equation} 
        for arbitrary $j \in \{0,\ldots,M-1\}$. Here $|F_j\rang$ is the computational basis state which is the binary representation of $F_j$. This step can be broken down into two. 
        
        \begin{enumerate}
            \item First, we perform the transformation 
                \begin{equation}\label{e_amp}
                    \mathcal{E}^\text{amp}|j\rang |0\rang = |j\rang |\Psi_j\rang,
                \end{equation} 
            where $|\Psi_j\rang$ is a state with information regarding $F_j$ encoded in its amplitudes. We achieve this using the Swap test algorithm.
            \item We now perform the transformation
                \begin{equation} \label{e_dig}
                    \mathcal{E}^\text{dig}|j\rang |\Psi_j\rang = |j\rang |F_j\rang.
                \end{equation}
            This can be thought of as an analog to digital conversion as we are converting the fidelity information from amplitudes of $|\Psi_j\rang$ to a digital format $|F_j\rang$. We use a slightly modified version of the abs-QADC algorithm to achieve this.
            \end{enumerate}
        Thus, $\mathcal{F} = \mathcal{E}^\text{dig} \mathcal{E}^\text{amp}$.
        \item Consider two pairs of registers, \emph{index, fidelity; index$'$, fidelity$'$} initialised as $|j\rang_{\text{in}} |0\rang_{\text{fid}} |y\rang_{\text{in$'$}} |0\rang_{\text{fid$'$}}.$    
        
        Apply $\mathcal F$ (step \ref{briefstep1}) on each of the two pairs of registers.     
        \begin{equation}
            \xrightarrow{{ \mathcal{F}_{\text{in,fid}} \mathcal{F}_{\text{in}',\text{fid}'}}} \\|j\rang_{\text{in}} |F_j\rang_{\text{fid}} |y\rang_{\text{in$'$}} |F_y\rang_{\text{fid$'$}}.
        \end{equation}
        where $\mathcal{F}_{\text{in,fid}}$ denotes the gate $\mathcal{F}$ applied on the index and fidelity registers.
        
        \item Now that we have our information in digital format, we may use a series of classical gates to realise the function $f_{y,A}~\eqref{eq:fya}$. Note that any classical operation can be simulated in a quantum setting using Toffoli gates (refer page 29 of~\cite{Nielsen2011}). Let $\mathcal C$ denote the operator achieving~\eqref{eq:fya}. Uncomputing the irrelevant registers, we obtain
        \begin{equation}
            \xrightarrow{\mathcal C, \text{uncompute}} |j\rang |f_{y,A}(j)\rang.
        \end{equation}
        For the sake of brevity, we do not further expand upon the form of $\mathcal C$, just like with other gates, here.         
        
    \end{enumerate}
    We now provide an explicit construction of the oracle $\mathcal O_{y,A}$.
      
    \begin{figure}
                \begin{center}
                    \includegraphics[scale=0.97, page=2]{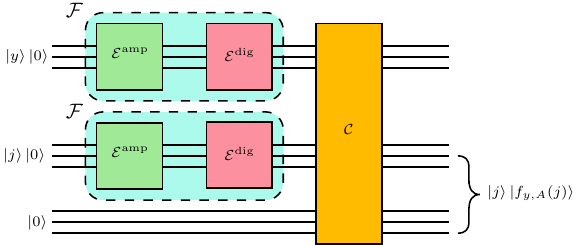}
                    \caption{A breakdown of the oracle $\mathcal O_{y,A}$. In two pairs of registers correspoding to index $j$ and $y$, apply $\mathcal F$ separately to obtain digital encoding of $F_j$ and $F_y$. Using classical circuits (through Toffoli gates) compare $F_j$ and $F_y$, and flip an ancilla qubit from $|0\rang$ to $|1\rang$ if $F_j>F_y$. Uncomputing irrelevant registers, we obtain $|j\rang |f_{y,A}(j)\rang$. Note that in the explicit construction of the oracle, there is an uncomputation procedure of one $\mathcal{F}$ coming in between the required classical operations. But for simplicity of explaining the circuit, we have avoided it in this diagram. }
                    \label{fig:OyaBreakdown}
            \end{center}
    \end{figure}
        
    \subsection{Constructing the oracle $\mathcal O_{y,A}$}
    We begin with the assumption that we are provided with efficient circuits of \textit{state preparation} oracles $\mathcal V, \mathcal W$ of the form
    \begin{gather}
        \mathcal V |0^n\rang  = |\psi\rang,\\
        \mathcal W|j\rang |0^n\rang =|j\rang|\phi_j \rang,
    \end{gather}
    for $j \in \{0, \ldots, M-1\}$. That is, we do not require the classical description of the test state. Instead we require the circuits. Efficient implementation of oracles like $ \mathcal{W}$ is discussed in~\cite{Kerenidis2016}. Assuming such oracles are provided, we now describe the construction of the oracle $\mathcal O_{y,A}$.

    \begin{enumerate}
        \item 
            \label{step1}
            Initialise four registers named \textit{index, train, test}, $B$ of sizes $m, n, n, 1$ respectively, where $n= \log N$ and $m = \log M$. 
            \begin{equation}
                |j\rang_{\text{in}}|0^{\otimes n}\rang_{\text{tr}}|0^{\otimes n}\rang_\text{tst}|0\rang_B.
            \end{equation}

        \item \label{fid_step2} 
            Apply $\mathcal W$  on train register 
            \begin{equation}
                     \xrightarrow{{\mathcal W _\text{in,tr}   }}  |j\rang_\text{in} |\phi_j\rang_{\text{tr}}|0\rang_\text{tst}|0\rang_B  .   
            \end{equation}
        
        \item  \label{V_oracle}    
            Now apply $\mathcal V$ on test register to obtain
            \begin{equation}
                \xrightarrow{{\mathcal V _\text{tst}}}  |j\rang_\text{in} |\phi_j\rang_{\text{tr}}|\psi\rang_\text{tst}|0\rang_B.
            \end{equation}
        
        \item \label{step4} 
            Apply the swap test circuit (sans measurement) between train register and test register with $B$ as the control qubit. The state is then
            \begin{equation}
                \label{swapteststate}
                \begin{split}
                    \xrightarrow{{ \text{SwapTest}}}\\
                    \frac{1}{2} |j\rang_\text{in}\bigg[ &\Big(|\phi_j\rang_{\text{tr}}|\psi\rang_\text{tst} + |\psi\rang_{\text{tr}}|\phi_j\rang_\text{tst}\Big)|0\rang_B  \\+ 
                    &\Big(|\phi_j\rang_{\text{tr}}|\psi\rang_\text{tst} - |\psi\rang_{\text{tr}}|\phi_j\rang_\text{tst}\Big)|1\rang_B
                    \bigg] \\\equiv  & |j\rang_{\text{in}}|\Psi_j\rang_{\text{tr,tst}, B},
                \end{split}
            \end{equation}
            where we have defined 
            \begin{equation}
            \begin{split}
                |\Psi_j\rang = \frac12  \Big[&\Big(|\phi_j\rang_{\text{tr}}|\psi\rang_\text{tst} + |\psi\rang_{\text{tr}}|\phi_j\rang_\text{tst}\Big)|0\rang_B  \\+ 
                 &\Big(|\phi_j\rang_{\text{tr}}|\psi\rang_\text{tst} - |\psi\rang_{\text{tr}}|\phi_j\rang_\text{tst}\Big)|1\rang_B\Big].
            \end{split}
            \end{equation}
             
            Define $U$ to be the combined unitary transformations of steps \ref{V_oracle} and \ref{step4} (refer Fig.~\ref{fig:U_circuit}). If one now measures the register $B$, one would see the probabilities as    
            \begin{gather}
                \text{Pr}(B = 0) = \frac{1+F_j}{2}, \\ \text{Pr}( B = 1) = \frac{1-F_j}{2}.
            \end{gather}
        
        The information regarding fidelity is now encoded in the amplitudes. Therefore gates from steps \ref{fid_step2}-\ref{step4} makes up the $\mathcal{E}^\text{amp}$ operator given in Eq.~\eqref{e_amp}.  We must now convert it into a $\lq$digital' format which can be further utilised.

            \begin{figure}
                \begin{center} 
                    \includegraphics[scale = 0.85, page=2]{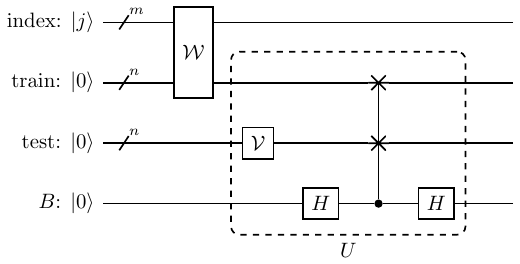}
                    \caption{Circuit for steps \ref{step1}-\ref{step4}.} \label{fig:U_circuit}
                \end{center}
            \end{figure}

            \begin{figure}
                    \begin{center} 
                        \includegraphics[scale = 0.8, page=6]{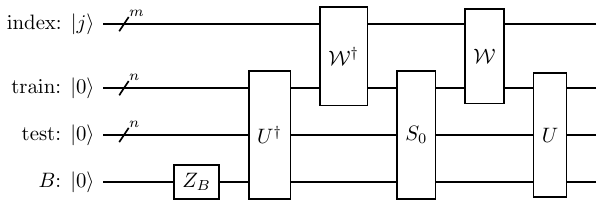}
                        \caption{Constructing the operator $G$ as defined in Eq.~\eqref{eq:G}.} \label{fig:Gcircuit}
                    \end{center}
            \end{figure}

            \begin{figure*}
                \begin{center}
                    \includegraphics[width=\textwidth, page=1]{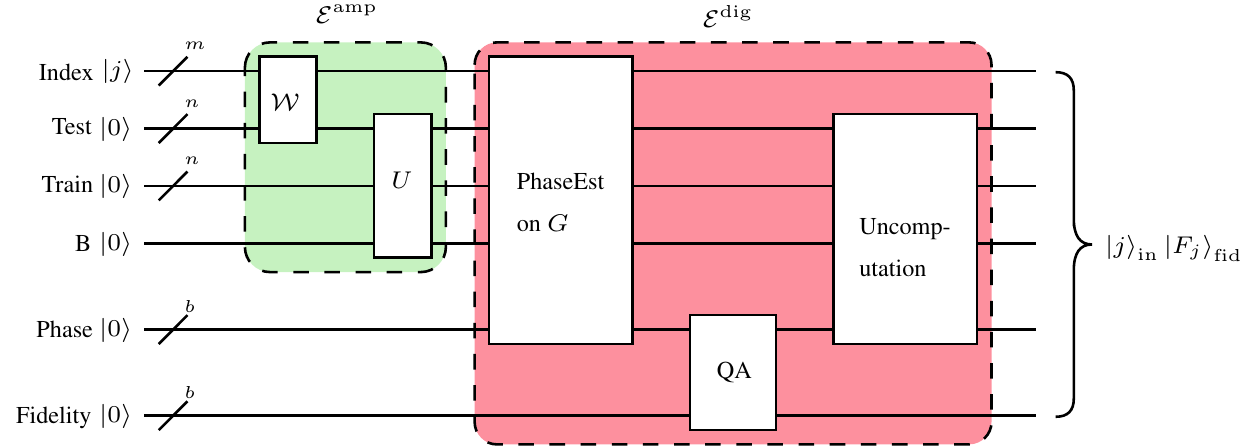}
                    \caption{Detailed construction of the operator $\mathcal F$ defined in Eq.~\eqref{F_j_oper}. We first use $\mathcal E^{\text{amp}}$ to encode the fidelity in the amplitudes. See Fig. (\ref{fig:U_circuit}) for details on $U$. Having the fidelity $F_j$ encoded as amplitudes, we want to convert them into a digital format. To this end we use $\mathcal E^{\text{dig}}$, which comprises of phase estimation on the operator $G$ (Eq.~\eqref{eq:G}) (refer Fig. \ref{fig:PhaseEst} for an explicit construction), which returns a state that must undergo further arithmetics (realised using quantum arithmetics) before it has a digital representation of $F_j$ stored in the \textit{Fidelity} register. We represent these circuits as QA. Finally, uncomputing every register except the Index and Fidelity registers, we have the required state.} 
                    \label{fig:Fcircuit}
                \end{center}
            \end{figure*}

        \item 
            \label{step5}
            To this end, construct a gate
            \begin{equation}\label{eq:G}
                G = U_{\text{tr,tst},B} \mathcal W_{\text{in,tr}}S{_{0_{\text{tr,tst,B}}}} \mathcal W^\dag_{\text{in,tr}} U^\dag_{\text{tr,tst},B} Z_B , 
            \end{equation}
            where $Z_B$ denotes the application of the $Z$ gate on register $B$ and $S_0 = \mathds{1} - 2 |0\rang \lang 0|$ (refer Fig.~\ref{fig:Gcircuit}). This operator can be seen as the operator $G$ used in the abs-QADC algorithm~\cite{Mitarai2019}, with the CNOT gates replaced by the train data preparation oracle $\mathcal{W}$. The action of $G$ on the current state can be written as controlled action of operators $G_j$ (refer appendix \ref{G_and_G_k}):
            \begin{equation}
            \begin{split}
                G  |j\rang_\text{in}|\Psi_j\rang_{\text{tr,tst}, B} =  |j\rang_\text{in}\Big(G_j|\Psi_j\rang_{\text{tr,tst}, B}\Big), 
            \end{split}
            \end{equation}
            
            where
            \begin{align}
                G_j &= U_{\text{tr,tst},B}S_j U_{\text{tr,tst},B}^\dag Z_B, \\ 
                S_j &= \mathds{1}  - 2\Big(|\phi_j\rang\lang \phi_j|_{\text{tr}} \otimes |0\rang \lang 0|_{\text{tst},B } \Big).  
            \end{align}

        \item
                 $|\Psi_j\rang_{\text{tr,tst}, B}$ can be decomposed into two eigenstates of $G_j$, namely $|\Psi_{j+}\rang$ and $|\Psi_{j-}\rang$, corresponding to the eigenvalues $e^{\pm i 2\pi \theta_j}$, respectively. Here, $\sin (\pi\theta_j) = \sqrt{\frac{1}{2}(1 +F_j)}$ and $\theta \in [1/4, 1/2)$ (refer Appendix \ref{app:G_Eig}). The decomposition is given as
            \begin{equation}
                |\Psi_j \rang = \frac{-i}{\sqrt{2}} (e^{i\pi \theta_j} |\Psi_{j+}\rang - e^{-i\pi\theta_j} |\Psi_{j-}\rang),
            \end{equation}
        
        \item
            \label{step6}
            We now have an operator $G$ which has the fidelity value $F_j$ stored in its eigenvalues. To get a $b$-bit binary representation of $\theta_j$, we now run the quantum phase estimation algorithm (refer Appendix \ref{app:qpe} for more details on quantum phase estimation) on $G$. To this end, we bring the \textit{phase register} containing $b$ qubits and run the phase estimation algorithm:
            \begin{equation}                
                \begin{aligned}
                \xrightarrow{\text{PhaseEst.}}
                &\frac{-i}{\sqrt{2}}  |j\rang_{\text{in}} \Big[
                e^{i\pi \theta_j} |\theta_j\rang_{\text{ph}}  |\Psi_{j+}\rang_{\text{tr,tst,} B} 
                \\ &- e^{-i\pi\theta_j} |1-\theta_j\rang _{\text{ph} }|\Psi_{j-}\rang_{\text{tr,tst}, B}\Big]\\
                &\equiv  |j\rang_\text{in} |\Psi_{j,\text{AE}}\rang_{\text{\text{ph,tr,tst,}}B}. 
                \end{aligned}
                \label{Eq:est}
            \end{equation}
            where we have defined the combined state of all registers except index register after estimation to be
            \begin{equation}
            \begin{split}
                |\Psi_{j_{\text{AE}}}\rang_{\text{ph,tr,tst,}B} =
                \frac{-i}{\sqrt 2}\Big(e^{i\pi \theta_j}|\theta_j\rang_{\text{ph}}  &|\Psi_{j+}\rang_{\text{tr,tst,} B} 
                   \\ - e^{-i\pi\theta_j} |1-\theta_j\rang_{\text{ph} }|\Psi_{j-}&\rang_{\text{tr,tst}, B}\Big).
            \end{split}
            \end{equation}
            Here, $|\theta_j \rang_\text{ph}$ and $|1- \theta_j\rang_\text{ph}$ are $b$-qubit states storing $b$-bit binary representation of $\theta_j$ and $1- \theta_j$ respectively.

            \item
            Introducing a separate register, named \textit{fid}, compute $F_j = 2 \sin^2(\pi\theta_j) - 1$ using quantum arithmetic (Appendix \ref{qa}). Note that $\sin (\pi\theta_j) = \sin(\pi(1-\theta_j))$, and $F_j$ is uniquely recovered. Then our total state is 
            \begin{equation}
            \begin{split}
                \xrightarrow {\text{ q. arithmetics }}
                |j\rang_\text{in} |F_j\rang_\text{fid} |\Psi_{j, \text{AE}}\rang_{\text{ph,tr,tst,}B}.
            \end{split}
            \end{equation}
            
            \item \label{fid_step9} Uncompute everything in registers phase, train, test and $B$ to get
            \begin{equation}
                \xrightarrow {\text{uncompute ph, tr, tst, } B} |j\rang_\text{in} |F_j\rang_\text{fid}.
            \end{equation} 
            Now, we have successfully converted the fidelity values from amplitudes to digital format. Therefore steps \ref{step5}-\ref{fid_step9} makes up the operator $\mathcal{E}^\text{dig}$ given in Eq.~\eqref{e_dig}. \ref{fid_step2}-\ref{fid_step9} gives the construction of the gate $\mathcal F$ given in Eq.~\eqref{F_j_oper}. We now have an operator capable of the transformation
            $    |j\rang |0\rang \xrightarrow {\mathcal F} |j\rang |F_j\rang$ 
            for arbitrary index $j$.
            \item On separate registers, named index$'$ and fidelity$'$, initialised as $|y\rang_{\text{in}'}|0\rang_{\text{fid}'} $, apply $\mathcal F$, to obtain
            \begin{equation}
                \xrightarrow {\mathcal F} |j\rang_\text{in} |F_j\rang_\text{fid} |y\rang_{\text{in}'} |F_y\rang_{\text{fid}'}.   
            \end{equation}

        \item 
            Add an extra qubit $Q_1$ and apply the classical comparison gate (refer Appendix \ref{app:bs_comp} for more details)
        \begin{equation}
            J|a\rang |b\rang |0\rang = 
            \begin{cases}
                |a\rang |b\rang |1\rang &: a >b, \\
                |a\rang |b\rang |0\rang &: a \leq b,
            \end{cases}
        \end{equation} \label{jgate}
        on registers fid and fid$'$ to get the state
        \begin{equation}
            \begin{split}
                \xrightarrow{J}|j\rang_\text{in} |F_j\rang_\text{fid} &|y\rang_\text{in'} |F_y\rang_{\text{fid$'$}} |g(j)\rang_{Q_1}
            \end{split}
        \end{equation}
        where
        \begin{equation}
            g(j) = 
            \begin{cases}
                1 &: F_j >F_y, \\
                0 &: F_j \leq F_y,
            \end{cases}
        \end{equation}
        The qubit $Q_1$ will mark all indices $j$ such that $F_j > F_y$.
        \item 
            Uncompute the registers in$'$ and fid$'$ to obtain the state
            \begin{equation}
                \begin{split}
                    \xrightarrow{\text{uncompute in$'$, fid$'$}}|j\rang_\text{in} |F_j\rang_\text{fid}|g(j)\rang_{Q_1}.
                \end{split}
            \end{equation} \label{saving_m_qubits_fid}            
        \item 
            Add an extra qubit $Q_2$ and for every $i_l \in A$, apply the gate $D^{(i_l)}$ of the form 
            \begin{equation}\label{M_gate_fid}
            D ^ {(i_l)}|j\rang |0\rang = 
            \begin{cases}
                |j\rang |1\rang &: j = i_l,\\
                |j\rang |0\rang &: j \neq i_l. 
            \end{cases}
        \end{equation} \label{dgate}
            
            on registers index and $Q_2$ to get the state
            \begin{equation}
            \begin{split}    
                \xrightarrow{\left(D^{(i_1)} \cdots D^{(i_k)}\right)_{\text{in,}Q_1}}&\\ |j\rang_\text{in} |F_j\rang_\text{fid} &|g(j)\rang_{Q_1} |\chi_A(j)\rang_{Q_2},
            \end{split}
            \end{equation}
            where $\chi_A(j) = 1$ if $j \in A$ and $0$ otherwise, is the indicator function of the set $A$. That is, the sequence of operators $D^{(i_1)}\cdots D^{(i_k)}$ marks all the indices that are already in the threshold index set $A$ as we would like to avoid these indices so as to not have repetition in our top $k$ neighbors. These gates can be realized using classical gates (refer Appendix \ref{app:bs_comp} for more details) by preparing the state $|i_l \rangle$ in an additional $m$-qubit register, which we uncomputed and recycled from step \ref{saving_m_qubits_fid}, and using comparison gates.
        \item 
            Add an extra qubit $Q_3$. Then apply an $X$ gate on $Q_2$ and a Toffoli gate with controls $Q_1, Q_2$ and target $Q_3$. This results in the state
            \begin{equation}
                \begin{split}
                    \xrightarrow{X, \text{Toffoli}}& \\|j\rang_\text{in} &|F_j\rang_\text{fid} |g(j)\rang_{Q_1} |\chi_A(j)\rang_{Q_2} |f_{y,A}(j)\rang_{Q_3}
                \end{split}
            \end{equation}
            Note that ultimately we are trying to construct an oracle capable of marking indices $j$ which have $g(j) = 1$ as well as $\chi_A(j) = 0$. An $X$ gate combined with a Toffoli gate will flip the target qubit if one of the input qubits is $0$ and the other is $1$. 
            \item
            Uncomputing every register except index and $Q_3$, we have
            \begin{equation}
                \xrightarrow {\text{uncompute}}|j\rang_{\text{in}}|f_{y, A}(j)\rang_{Q_3}.
            \end{equation}
            Since the construction of this circuit does not depend on $j$, we now have an operator that does the aforementioned transformation. 
            \begin{equation}
                \mathcal O_{y, A}|j\rang|0\rang = |j\rang|f_{y,A}(j)\rang.
            \end{equation}
    \end{enumerate}
    This completes the construction of the oracle $\mathcal O_{y,A}$~\eqref{OracleOFid}. We may now use this oracle in the $k$ maxima finding algorithm to find the $k$ nearest neighbors of a test state based on fidelity.

\section{Quantum $k$ Nearest neighbors algorithm using dot product}\label{sec:qknndot}    
    We briefly discuss how one can construct a  Q$k$NN algorithm that utilises dot product $X(u,v) \equiv \lang u|v\rang$ instead of fidelity. For real-valued vectors, which is usually the case with real-world practical applications, dot product is more useful. Note that for  Q$k$NN using fidelity, we are performing Swap test between the test state and all train states in superposition to analog-encode the fidelity information $F_j$, and we then use abs-QADC algorithm to digitise this information. For  Q$k$NN using dot product, we replace Swap test with Hadamard test (refer Appendix \ref{app:HadamardTest}), i.e., we perform Hadamard test between the test state and all the train states in superposition to analog-encode the dot product information $X_j$, and then use a similar modification to the real-QADC algorithm to digitise this information. Note that even for complex-valued vectors $|\psi\rang, |\phi\rang$, real-QADC returns the real part of the inner product $\text{Re} \lang \psi|\phi\rang$, and thereby can be of use if the situation agrees. 
    
    The steps of this procedure are similar to  Q$k$NN using fidelity. Therefore we refrain from presenting it here and instead present it in full detail in Appendix \ref{app:QKNNdot}.

\section{Complexity analysis}\label{sec:comp}
    The fidelity based  Q$k$NN requires $O(\sqrt{kM})$ calls to the oracle $\mathcal{O}_{y,A}$. The oracle contains two uses of the $\mathcal{F}$ circuit, which is a slight modification of the abs-QADC algorithm. Therefore the complexity of executing the oracle is similar to the abs-QADC algorithm. For a precision parameter $\epsilon = 2 ^ {-b}$, the circuit of the oracle contains $O(1/\epsilon)$ controlled $\mathcal{V}$ and $\mathcal{W}$ gates and $O((\log^2N)/\epsilon)$ single and 2 qubit gates. So, one can execute the whole fidelity based  Q$k$NN procedure with $O(\sqrt{kM})/\epsilon)$ calls to the data preparation oracles $\mathcal{V}$ and $\mathcal{W}$ and $O(\sqrt{kM}(\log^2N)/\epsilon)$ single and 2 qubit gates, $O(k\sqrt{kM})$ $D^{(i_l)}$ gates, each of which can be realized in $O(\log (M))$ with an extra register of $\log M$ qubits and $O(\sqrt{kM})$ $J$ gates, each of which can be realized in $O(b)$. 
    
     When $ N >> 1/ \epsilon$, Q$k$NN algorithm will be a better choice than other $k$NN algorithms that require the description of the test state since reading the description of the state will have complexity $O(N)$, regardless of how their complexity is related to $\epsilon$. Another approach that can bypass the process of reading the description of the state is by estimating the fidelity through sampling from measurements. For example, one can carry out swap test between the test state and each train state separately, estimate fidelity in each case, and then find the $k$ nearest neighbours. Now, in this method, since the fidelity values are estimated by sampling, to achieve an error bound of $\epsilon $, the test state preparation circuit $\mathcal V$ has to be executed $O({1 / \epsilon ^ 2})$ times. Then, the method will have a query complexity which is at least $O(M/ \epsilon^2)$. The proposed Q$k$NN has a query complexity of $O(\sqrt{kM} / \epsilon)$. So, in this scenario also, Q$k$NN has a potential advantage. 
    
    The Q$k$NN algorithm can be executed in $O(\log M) + O(\log N) + O(\log 1/ \epsilon)$ qubits. To perform $\mathcal{F}$ once, we need $\log M + b$ qubits and another $2\log N + b + 1$ qubits which are uncomputed at step \ref{fid_step9}. Considering the fact that the uncomputed qubits can be recycled, applying $\mathcal{F}$ on two separate registers will require $2(\log M + b) + 2\log N + b + 1$ qubits. The classical operations would require $3$ qubits as well as $ O(\log M) $ qubits for the $D ^ {(i_l)}$ gates and $ O(b)$ qubits for the $J$ gates. Let $k_J$ be the number of qubits required for the $J$ gates and let $k_D$ be the number of qubits required for the $D ^ {(i_l)}$ gates. Since the $2\log N + b + 1$ qubits can be recycled and after using the $J$ gates, the states prepared in registers in$'$, fid$'$ can be uncomputed, to execute the Q$k$NN algorithm using fidelity measure, we require $ 2 \log M + 2b + k_J + 1 + \text{max}(k_D + 2 - \log M - b, 2 \log N + b - k_J)$. For example, to carry out Q$k$NN for $30$ qubit states (dimension $\approx 1$ billion) with $2 ^ {10}$ train states, we would require around $84 + 3b$ qubits, along with any ancilliary qubits required $k_D$ and $k_J$.
    
    The dot product based  Q$k$NN has the same query complexity as the fidleity based  Q$k$NN. To perform this operation, one would require $ 2 \log M + 2b + k_J + 1 + \text{max}(k_D + 2 - \log M - b, \log N + b - k_J)$ qubits.

\section{Applications}\label{sec:app}
    We present two scenarios where our quantum $k$NN algorithm can be applied. In both scenarios, we utilize the fact that we can classify an unknown test state. 
    
    \subsection{Entanglement classification}
    
    \subsubsection{Entanglement classes}
         In this section, we discuss the entanglement classes in pure $n$-partite quantum states. For simplicity, we restrict ourselves to $n$-qubit systems. We begin with $n=2$ case. A pure two-qubit quantum state $|\Phi\rangle$ is called separable or product state if and only if it can be written as a tensor product of two pure states corresponding to individual subsystems,
        \begin{equation}\label{eq:sep_states}
            |{\Phi}\rang = |{\phi_1}\rang \otimes |{\phi_2}\rang.
        \end{equation} 

        If the state $|{\Phi}\rang$ is not of the form \eqref{eq:sep_states} then its an entangled state. A pure $n$-qubit quantum state $|\Psi\rang$ is separable only  if it can be written as the tensor product of $n$ quantum states as 
        \begin{align}
           |{\Psi}\rang = |{\psi_1}\rang\otimes \cdots \otimes |{\psi_n}\rang.
        \end{align}
        Such states are also called $n$-separable states~\cite{Horodecki2009}. Equivalently, a pure state $|\Psi\rang$ is an $n$-separable state if it is separable across all the possible bipartitions of the $n$ qubits. If this condition is violated, then the state is no longer $n$-separable. Some states can be entangled in certain bipartitions and separable in others. Some states are entangled across all bipartitions. This motivates a classification of $n$-partite quantum states on the basis of entanglement. For two-qubit systems, there are only two classes -- separable and entangled states. 
        
        \label{para:3qubit}For three-qubit systems, we have more. Let A, B, and C represent the three qubits, and let us use A-B to denote that subsystems A and B are separable and AB to denote that subsystems A and B are entangled. Then the entanglement classes can be written as \{A-B-C,~AB-C,~A-BC,~AC-B,~ABC\}. Note that we do not distinguish between W states and GHZ states defined in~\cite{Dur2000} and keep them in the same class ABC.
        
        The same classification of the entanglement can be extended to  $n$ number of qubits. The question that is relevant to us is the following: given a circuit that prepares an $n$-qubit arbitrary state, is there a way to label it according to its entanglement class. We show that the classical $k$NN algorithm can classify these states for $n=2, 3$ with high accuracy. Furthermore, similar accuracy can be achieved by our Q$k$NN algorithm without the classical description of the given quantum state, establishing the advantage of Q$k$NN over the other $k$NN algorithms in settings where the input is a circuit which prepares the test state, rather than the description of the state itself.
        
    \subsubsection{Simulation results}
    
    We present the results of simulation of estimating the entanglement class of a test state $|\psi\rang$ using classical $k$NN in Table \ref{tab:classicalsimulations}~\footnote{The simulation code can be found at \url{https://github.com/afradnyf/QKNN}.}. We use this numerical experiment to demonstrate that quantum states in Hilbert space have a nearest-neighbor structure when it comes to entanglement. That is, the closer the states are, the similar their entanglement is. 
    
    For two-qubit states, we demonstrate this for both separable (entanglement entropy $= 0$) vs entangled (entanglement entropy $\neq 0$) and separable vs maximally entangled (entanglement entropy $= 1$). We also demonstrate it for three-qubit states with a classification among the five different classes mentioned above.

    \begin{table*}
        \center
        \begin{tabular}{|c|c|c|c|} 
         \hline
         No. of Qubits & No. of classes & Entanglement classes & Accuracy \\ 
         \hline
         2 & 2 & Separable, Entangled & 99\% \\ 
         \hline
         2 & 2 & Separable, Maximally entangled & 100\% \\
         \hline
         3 & 5 & 1-2-3, 12-3, 1-23, 13-2, 123 & 89\%  \\
         \hline
        \end{tabular} 
        \caption{Entanglement classification using classical $k$NN classifier. Cardinality of the set of train states is $M =$ (number of classes)$\times$(class size). In each case, the total number of train states used for each class is $10 ^ 5$.}
        \label{tab:classicalsimulations}
    \end{table*}

    \subsection{Quantum state discrimination}
    Another application that we propose for  Q$k$NN is a problem analogous to quantum state discrimination~\cite{Bae2015}. The problem of quantum state discrimination is originally formulated for arbitrary (mixed) states and in terms of measurement. Consider a collection of states $\mathcal S = \{\rho_0, \ldots, \rho_{M-1}\}$ with associated probabilities $p = (p_0, \ldots, p_{M-1})$. A state is drawn from $S$ according to $p$ and is prepared. The aim is to find a measurement which maximizes the probability of correctly identifying the prepared state.  
    
    We are concerned with an analogous problem. Suppose we are given a circuit that produces an unknown (pure) state $|\psi\rang$. We are guaranteed that $|\psi\rang$ is one of the $M$ known (pure) states $\{|\phi_0\rang, \ldots, |\phi_{M-1}\rang\}$. We also assume that we are given circuit $\mathcal{W}$ to prepare the known states. The task is to correctly deduce the identity of $|\psi\rang$. 
    
    In such a scenario, one can run the  Q$k$NN algorithm with $k=1$ to obtain the $j$ such that $|\psi\rang = |\phi_j\rang$ with $ O(\sqrt{M})$ oracle calls, as the largest value fidelity can take is $1$ and $\text{F}(u,v) = 1$ if and only if $u=v$.

\section{Conclusion}\label{sec:conclusion}
    In this paper, we have presented a novel Q$k$NN algorithm, which is a quantum analog of the classical $k$NN algorithm. We use Swap test and generalizations of quantum analog to digital conversion algorithms to construct an oracle which enables us to reduce the problem of quantum $k$NN to an instance of quantum $k$ maxima finding algorithm. We assume that state preparations circuits are provided, and the algorithm uses fidelity as a similarity measure which is widely used in problems where the data is inherently quantum. The algorithm requires $O(\sqrt{kM})$ calls to these circuits to obtain the identities of the $k$ nearest neighbors of the test out of $M$ train states. Since the metric Bures distance is a monotonous function of fidelity, $k$NN done using fidelity is in agreement with $k$NN carried out using Bures distance. We also present a variant of the algorithm which uses dot product as distance measure. 
    
    An advantage of the proposed algorithm is its ability to classify quantum states without their explicit classical description in some basis. Instead, we require circuits capable of preparing these states.  Furthermore, while dealing with quantum data, the algorithm is able to classify without the requirement of quantum state tomography, which is essential if one were to use any other, quantum or classical, $k$NN algorithm which requires the classical description of states. Along with the calls to the state preparation circuits being $O(\sqrt{kM})$, all the other parts of the circuit are efficiently preparable. The number of qubits required is also poly-logarithmic in the dimension of the states involved and the number of train states. 
    
    As an example, we show the effectiveness of $k$NN method in identifying the type of entanglement in quantum states. For the problem of entanglement classification, preparing train states of different types of entanglement is much easier when we're working with their circuits rather than working with their classical descriptions. Furthermore, we discuss the applicability of the algorithm in an analogous version of quantum state discrimination. 
    
    A particular future direction regarding the applications of Q$k$NN is to study its capability in an analogous version of quantum gate discrimination~\cite{Chiribella2013}. This is due to the ability of Q$k$NN to work with quantum circuits which could be efficient representations of unitary matrices. The problem of identifying a test circuit among a finite set of train circuits can potentially be addressed using a column-wise Q$k$NN.

\section{Acknowledgements}
We thank Yuan Feng, Sanjiang Li, and Christopher Ferrie for fruitful discussions. SKG acknowledges the financial support from SERB-DST (File No. ECR/2017/002404) and Interdisciplinary Cyber Physical Systems(ICPS) programme of the Department of Science and Technology, India (Grant No.:DST/ICPS/QuST/Theme-1/2019/12). The quantum circuits were generated using the Quantikz package \cite{Kay2018} and Mathcha~\cite{Mathcha}.

\bibliographystyle{plain}

\appendix
\onecolumn

\section{Quantum Analog-Digital Conversion algorithms}\label{app:QADC}
Mitarai et al. describes a set of algorithms~\cite{Mitarai2019} to carry out analog-digital conversions within a quantum circuit. Let $U|0\rangle = \sum_{i=0}^{d-1} c_i |i\rang$ be an arbitrary quantum state.
    
    \begin{definition}{abs-QADC}
        ~\cite{Mitarai2019} Let $\widetilde{r_j}$ denote the m-bit string $\widetilde{r_j} ^ {1}, \widetilde{r_j} ^ {2}, \dots , \widetilde{r_j} ^ {m}$ that best approximates $|c_j|$ by $\sum \limits_{k = 1}^{m} \widetilde{r_j} ^ {k} 2 ^ {-k}$. An m-bit abs-QADC operation transforms analog-encoded state $\sum \limits_{j = 1} ^ {N} c_j |j\rang |0\rang ^ {\otimes m} $ to $\frac{1}{\sqrt{N}} \sum \limits_{j = 1} ^ {N} |j\rangle |\widetilde{r_j} \rangle.$

    \end{definition}
    
    \begin{definition}{real-QADC}
        ~\cite{Mitarai2019} Let $\widetilde{x_j}$ denote the m-bit string $\widetilde{x_j} ^ {1}, \widetilde{x_j} ^ {2}, \dots , \widetilde{x_j} ^ {m}$ that best approximates the real part of $c_j$ by $\sum \limits_{k = 1}^{m} \widetilde{x_j} ^ {k} 2 ^ {-k}$. An m-bit real-QADC operation transforms analog-encoded state $\sum \limits_{j = 1} ^ {N} c_j |j\rangle |0\rangle ^ {\otimes m} $ to $\frac{1}{\sqrt{N}} \sum \limits_{j = 1} ^ {N} |j\rang |\widetilde{x_j}\rang.$
    \end{definition}

    \begin{theorem}{abs-QADC}
        ~\cite{Mitarai2019} There exists an m-bit abs-QADC algorithm that runs using $O(1/\epsilon)$ controlled-$U$ gates and $O((log^2N)/\epsilon)$ single and two qubit gates with output state fidelity $(1 - O(poly(\epsilon)))$, where $\epsilon = 2^{-m}$.
    \end{theorem}

    \begin{theorem}{real-QADC}
        ~\cite{Mitarai2019} There exists an m-bit real-QADC algorithm that runs using $O(1/\epsilon)$ controlled-$U$ gates and $O((log^2N)/\epsilon)$ single and two qubit gates with output state fidelity $(1 - O(poly(\epsilon)))$, where $\epsilon = 2^{-m}$.
    \end{theorem}

    \begin{figure*}
            \begin{center}
                \includegraphics[page=5]{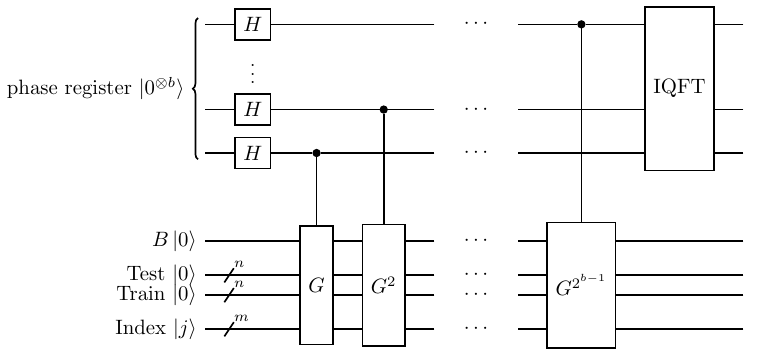}
                \caption{Quantum phase estimation on an operator $G$. This is the innards of the $\lq$PhaseEst on $G$' operator in Fig.~\ref{fig:Fcircuit}.}
                \label{fig:PhaseEst}
            \end{center}
        \end{figure*}

\section{Quantum Phase Estimation} \label{app:qpe}

    Quantum phase estimation is a quantum procedure that can be used to estimate the phase of the eigenvalue of a given eigenvector of a unitary operator. It relies on the quantum Fourier transform and is the engine behind some of the most popular quantum algorithms such as the Shor's algorithm for factoring.
    \begin{theorem}
        ~\cite{Cleve1998} Let $U$ be a unitary operator acting on M-qubit Hilbert space with eigenstates $\{ |\psi_j \rang \}_{j = 1} ^ {2 ^ M}$ and corresponding eigenvalues $\{ e ^ {2 \pi i \phi_j}\}_{j = 1} ^ {2 ^ M}$ where $\phi_j \in [0, 1)$. Let $\epsilon = 2 ^ {-m}$ for some positive integer $m$. There exists a quantum algorithm, which consists of $O(1/ \epsilon)$ controlled-$U$ calls and $O(log^2(1/ \epsilon))$ single and two-qubit gates, that performs transformation $\sum \limits_{j = 1} ^ {2 ^ M} a_j |\psi_j \rang |0 \rang ^ {\otimes m} \rightarrow | \psi_{PE} \rang = \sum \limits_{j = 1} ^ {2 ^ M} a_j |\psi_j \rang | \tilde{\mathbf{\phi}}_J \rang$ where $| \tilde{\mathbf{\phi}}_J \rang$ denotes a bitstring $\tilde{\phi_J} ^ {(1)} \tilde{\phi_J} ^ {(2)} \dots \tilde{\phi_J} ^ {(m)}$ such that $\Big| \sum \limits_{k = 1} ^ {m} \tilde{\phi_J} ^ {(k)} 2 ^ {-k} - \phi_J \Big| \leq \epsilon$ for all $j$ with state fidelity at least $1 - \text{poly}(\epsilon)$.
    \end{theorem}
\section{Quantum arithmetics} \label{qa}
    Within a quantum circuit, one can always apply simple arithmetic functions such as additions, multiplication, trginometric functions, exponentiation, etc. This is explained using the following theorems.
    \begin{theorem}
        ~\cite{Perez2017} Let $\mathbf{a}, \mathbf{b}$ be m-bit strings. There exists a quantum algorithm that performs transformation $|\mathbf{a} \rangle |\mathbf{b} \rangle \rightarrow |\mathbf{a} \rangle |\mathbf{a + b} \rangle$ with \textit{O}(poly(m)) single and two qubit gates.
    \end{theorem}
    Using this quantum adder, one can construct a circuit capable of carrying out any basic function in a similar manner in a quantum circuit.
    \begin{theorem}
        ~\cite{Mitarai2019} Some basic functions such as inverse, trigonometric functions, square root, and inverse trigonometric functions can be calculated to accuracy $\epsilon$, that is, we can perform a transformation $|a\rangle |0\rangle \rightarrow |a\rangle |\tilde{\mathbf{f(a)}}\rangle$ such that $|\mathbf{f(a)} - \tilde{\mathbf{f(a)}}| \leq \epsilon$ where $f$ is the required function using ${O}(\text{poly}(\log(1/\epsilon)))$ quantum arithmetics.
    \end{theorem} 
\section{Hadamard Test}\label{app:HadamardTest}
        The Hadamard test is a quantum circuit that can be used to compute the real part or the imaginary part of the inner product between two quantum states $|u\rang, |v\rang \in \mathbb{C} ^ {n}$. Since, we are more interested in the real part of the inner product, we will be using and explaining that particular version of the Hadamard test.
        
        \begin{figure}
            \begin{center}
                \includegraphics[page=4]{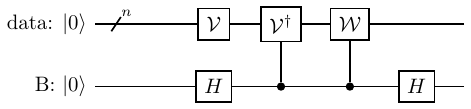}
                \caption{Circuit for Hadamard test.}
                \label{fig:hadamard}
            \end{center}
        \end{figure}
        
        Let $U |0\rang = |u\rang$ and $V |0\rang = |v\rang$. The aim is to use $U$ and $V$ in a controlled manner to construct the state 
        \begin{equation}
            \frac{1}{2} \Big[ |0\rang \Big( |u\rang + 
            |v\rang \Big) + |1\rang \Big( |u\rang - |v\rang \Big) \Big].
        \end{equation}
        The detailed circuit is shown in Fig.~\ref{fig:hadamard}. The real part of the inner product can be estimated from measuring the first qubit as
        \begin{equation}
          \text{Pr}(0)  = \frac{1}{2} + \frac{1}{2}\text{Re}(\langle u|v\rangle) \quad \text{and} \quad \text{Pr}(1)  = \frac{1}{2} - \frac{1}{2}\text{Re}(\langle u|v\rangle).
        \end{equation}
        The quantity $\text{Pr}(0)- \text{Pr}(1)$ gives us the desired real part of the inner product.

\section{Quantum $k$ Nearest neighbors algorithm using dot product}\label{app:QKNNdot}
     
    A more general distance measure used in $k$NN problems is the dot product. Let $\mathcal{H}$ be the $n$-qubit Hilbert space and let $|v\rang \in \mathcal H $ be the real valued test state whose label is to be determined. Let $\{|u_i\rangle : i \in \{0, \ldots, M-1\}\} \subset \mathcal H$ be a collection of $M$ known real valued train states whose labels are known to us. Let $X_i \equiv \text{X}(v, u_i) = \lang v|u_i \rang$ be the dot product between the test state $|v\rang$ and the $i^\text{th}$ train state $|u_i\rang$ and  
    \begin{equation}\label{setx}
        X = [X_0, \ldots, X_{M-1}]
    \end{equation} 
    be a table of length $M$ containing the dot product values with the test state $|v\rangle$ and all the train states $\{| u_i \rang\}$. 
    The general approach adopted here is very similar to fidelity based  Q$k$NN. 
    One can see that the problem of Q$k$NN using dot product is an instance of the quantum $k$ maxima finding algorithm on the set $X$ given in Eq.~\eqref{setx}. Similar to Q$k$NN using fidelity, the problem boils down to being able to perform quantum search on the Boolean function
    \begin{equation}
        f_{y,A}(j)  = 
        \begin{cases}
            1 &: X_j > X_y \text{ and } j \notin A,\\
            0 &: \text{otherwise}.                        
        \end{cases}
    \end{equation} \label{boolean_dp}
    That is, we should be able to prepare the oracle
    \begin{equation} \label{dotporacle}
        \mathcal O_{y, A}|j\rang|0\rang =
        \begin{cases}
        |j\rang|1\rang & : X_j > X_y \text{ and } j \notin A,\\
        |j\rang |0\rang &: \text{otherwise}.
        \end{cases}
    \end{equation}

    The assumptions here are that we are provided with \textit{state preparation} oracles $\mathcal V, \mathcal W$ of the form
        \begin{gather}
            |0^n\rang  \xrightarrow{\mathcal V} |v\rang,\\
            |j\rang |0^n\rang \xrightarrow{\mathcal W} |j\rang|u_j \rang,
        \end{gather}
        for $j \in \{0, \ldots, M-1\}$. The construction of the oracle is based on the real-QADC circuit from~\cite{Mitarai2019}. In that circuit, to compute the real values of the coordinates of the state, we apply the Hadamard test in superposition with the state and standard basis vectors. In the quantum $k$NN setting, we apply the Hadamard test in superposition with the test state and the train states. We now show the correctness of such a protocol and a method to build the required oracle $ \mathcal O_y$ using it. The explicit construction of the oracle is as follows:
        \begin{enumerate}
            \item 
                \label{step1_real}
                Initialise three registers named \textit{index, data}, $B$ of sizes $m, n, 1$ respectively, where $n= \log N$ and $m = \log M$
                \begin{equation}
                    |j\rang_{\text{in}}|0^{ n}\rang_{\text{data}}|0\rang_B.
                \end{equation}
            \item
                \label{step2_real}
                Apply $\mathcal V$ on the data register
                \begin{equation}
                    \xrightarrow{\mathcal{V}}|j\rang_{\text{in}}|v\rang_{\text{data}}|0\rang_B.
                \end{equation}
            \item
                \label{step3_real}
                Perform Hadamrd test to obtain the state
                \begin{equation}
                \begin{split}
                    \xrightarrow{\text{Hadamard Test}}&\frac{1}{2} |j\rang_{\text{in}}\Big[\Big( |v\rang_{\text{data}} + |u_j\rang_{\text{data}}\Big) |0\rang_B + 
                    \Big( |v\rang_{\text{data}} - |u_j\rang_{\text{data}} \Big) |1\rang_B \Big]\\ 
                    \equiv &|j\rang_\text{in}|\Psi_j\rang_{\text{data}, B},
                \end{split}
                \end{equation}
                where, 
                \begin{equation}
                    |\Psi_j\rang_{\text{data}, B} = \frac{1}{2} \Big[ \Big( |v\rang_{\text{data}} + |u_j\rang_{\text{data}}\Big) |0\rang_B + 
                    \Big( |v\rang_{\text{data}} - |u_j\rang_{\text{data}} \Big) |1\rang_B \Big].    
                \end{equation}
                Define $V$ to be the combined unitary transformations of steps \ref{step2_real} and \ref{step3_real}. If one now measures the $B$ register, one would see the probabilities as    
                \begin{equation}
                    \text{Pr}(B = 0) = \frac{1+X_j}{2} \quad \text{and} \quad \text{Pr}( B = 1) = \frac{1-X_j}{2}.
                \end{equation}
                The information regarding dot product is now encoded in the amplitudes. We must now convert it into a $\lq$digital' format which can be further utilised.

            \item 
                \label{step4_real}
                Construct a gate
                \begin{equation}
                    H = V_{\text{in,data,}B} \ \mathcal S_{0_{\text{data,}B}} \   V^\dag_{\text{in,data,}B} \ Z_B ,
                \end{equation}\label{eq:H}
                where $S_0 = \mathds{1} - 2 |0\rang \lang 0 |$. This $H$ gate can be seen as the $H$ gate in the real QADC algorithm~\cite{Mitarai2019}, with the controlled CNOT gate replaced by controlled $\mathcal{W}$. The action of $H$ on the current state can be written as controlled action of operators $H_j$ (refer Appendix \ref{actH}):
                \begin{equation}
                \begin{split}
                    H  |j\rang_\text{in}|\Psi_j\rang_{\text{data}, B} =  |j\rang_\text{in}\Big(H_j|\Psi_j\rang_{\text{data}, B}\Big), 
                \end{split}
                \end{equation}
                where
                \begin{equation}
                    H_j = \left(\mathds{1} - 2 {|\Psi_j \rang \lang \Psi_j |}_{\text{data,}B}\right) Z_B. 
                \end{equation}

            \item
                 $|\Psi_j\rang_{\text{tr,tst}, B}$ can be decomposed into two eigenstates of $H_j$, namely $|\Psi_{j+}\rang$ and $|\Psi_{j-}\rang$, corresponding to the eigenvalues $e^{\pm i 2\pi \theta_j}$, respectively. Here, $\sin (\pi\theta_j) = \sqrt{\frac{1}{2}(1 +X_j)}$ and $\theta \in [1/4, 1/2)$ (refer Appendix \ref{eigH}). The decomposition is given as
            \begin{equation}
                |\Psi_j \rang = \frac{-i}{\sqrt{2}} (e^{i\pi \theta_j} |\Psi_{j+}\rang - e^{-i\pi\theta_j} |\Psi_{j-}\rang),
            \end{equation}

        \item
            The operator $H$ has the dot product values $\{X_j\}$ stored in its eigenvalues. To get a $b$-bit binary representation of $\theta_j$, we now run the phase estimation algorithm on $H$. To this end, we bring the \textit{phase register} containing $b$ qubits and run the phase estimation algorithm:
            \begin{equation}                
                \begin{aligned}
                \xrightarrow{\text{PhaseEst.}}&
                \frac{-i}{\sqrt{2}}  |j\rang_{\text{in}} \Big[
                e^{i\pi \theta_j}|\theta_j\rang_{\text{ph}} |\Psi_{j+}\rang_{\text{data}, B} 
                - e^{-i\pi\theta_j} |1-\theta_j\rang _{\text{ph}}|\Psi_{j-}\rang_{\text{data}, B}\Big]\\
                &\equiv  |j\rang_\text{in} |\Psi_{j,\text{AE}}\rang_{\text{\text{ph,data,}}B}, 
                \end{aligned}        
            \end{equation}
            where we have defined the combined state of all registers except index register after estimation to be
            \begin{equation}
                |\Psi_{j,\text{AE}}\rang_{\text{\text{ph,data,}}B} =
                \frac{-i}{\sqrt{2}}\Big(e^{i\pi \theta_j}|\theta_j\rang_{\text{ph}} |\Psi_{j+}\rang_{\text{data,} B} 
                    - e^{-i\pi\theta_j} |1-\theta_j\rang _{\text{ph} }|\Psi_{j-}\rang_{\text{data}, B}\Big).
            \end{equation}
            Here, $|\theta_j \rang_\text{ph}$ and $|1- \theta_j\rang_\text{ph}$ are $b$-qubit states storing $b$-bit binary representation of $\theta_j$ and $1- \theta_j$ respectively.

        \item
            Introducing a register, \textit{dp}, compute $X_j = 2 \sin^2(\pi\theta_j) - 1$ using quantum arithmetic from Appendix \ref{qa}. Note that $\sin (\pi\theta_j) = \sin(\pi(1-\theta_j))$, and $X_j$ is uniquely recovered. Then our total state is 
            \begin{equation}
                \xrightarrow {\text{ quantum arithmetics }}
                |j\rang_\text{in} |X_j\rang_\text{dp} |\Psi_{j, \text{AE}}\rang_{\text{ph,data,}B}.
            \end{equation}
        \item
            \label{step8_real}
            Uncompute everything in registers ph,data and B to get
            \begin{equation}
                \xrightarrow {\text{uncompute ph, data, B}} |j\rang_\text{in} |X_j\rang_\text{dp}.
            \end{equation}
            We have successfully converted the dot product values from amplitudes to digital format.
        \item 
            Add more registers and apply $\mathcal X$, so that we have
            \begin{equation}
            \xrightarrow {\mathcal X} |j\rang_\text{in} |X_j\rang_\text{dp} |y\rang_\text{in$'$} |X_y\rang_{\text{dp$'$}}   
            \end{equation}
            
        \item 
            Add an extra qubit $Q_1$ and apply the gate $J$ defined in Eq.~\eqref{jgate} on registers dp and dp$'$ to get the state
        \begin{equation}
            \begin{split}
                \xrightarrow{J}|j\rang_\text{in} |X_j\rang_\text{dp} &|y\rang_\text{in$'$} |X_y\rang_{\text{dp$'$}} |g(j)\rang_{Q_1}
            \end{split}
        \end{equation}
        where
        \begin{equation}
            g(j) = 
            \begin{cases}
                1 &: X_j >X_y, \\
                0 &: X_j \leq X_y,
            \end{cases}
        \end{equation}
        \item 
            Uncompute the registers in$'$ and dp$'$ to obtain the state
            \begin{equation}
                \begin{split}
                    \xrightarrow{\text{uncompute in$'$, dp$'$}}|j\rang_\text{in} |X_j\rang_\text{dp}|g(j)\rang_{Q_1}.
                \end{split}
            \end{equation} \label{saving_m_qubits_dp}
        \item 
            Add an extra qubit $Q_2$ and for every $i_l \in A$, apply the gate $D^{(i_l)}$ defined in Eq.~\eqref{dgate} on registers index and $Q_2$ to get the state
            \begin{equation}
            \begin{split}    
                \xrightarrow{\left(D^{(i_1)} \cdots D^{(i_k)}\right)_{\text{in,}Q_1}} |j\rang_\text{in} |X_j\rang_\text{dp} |g(j)\rang_{Q_1} |\chi_A(j)\rang_{Q_2},
            \end{split}
            \end{equation}
            where $\chi_A(j) = 1$ if $j \in A$ and $0$ otherwise, is the indicator function of the set $A$. 
            
        \item 
            Add an extra qubit $Q_3$. Then apply an $X$ gate on $Q_2$ and a Toffoli gate with controls $Q_1, Q_2$ and target $Q_3$. This results in the state
            \begin{equation}
                \begin{split}
                    \xrightarrow{X, \text{Toffoli}}|j\rang_\text{in} |X_j\rang_\text{dp} |g(j)\rang_{Q_1} |\chi_A(j)\rang_{Q_2} |f_{y,A}(j)\rang_{Q_3},
                \end{split}
            \end{equation}
            where $f_{y,A}$ is the Boolean function defined in Eq.~\eqref{boolean_dp}.
        \item
            Uncomputing every register except index and $Q_3$, we have
            \begin{equation}
                \xrightarrow {\text{uncompute}}|j\rang_{\text{in}}|f_{y, A}(j)\rang_{Q_3},
            \end{equation}
            which is the required oracle
            \begin{equation}
                \mathcal O_{y, A}|j\rang|0\rang = |j\rang|f_{y,A}(j)\rang.
            \end{equation}
    \end{enumerate}
    This completes the construction of the oracle given in~\eqref{dotporacle}. We may now use this oracle in the $k$ maxima finding algorithm to find the $k$ nearest neighbors of a test state based on dot product.

\section{Action of $G$ as controlled $G_j$ for the Fidelity based  Q$k$NN}
\label{G_and_G_k}
    Recall that
    \begin{gather}
        G = U_{\text{tr,tst},B} \mathcal W_{\text{in,tr}}S_{0_{\text{tr,tst,B}}} \mathcal W^\dag_{\text{in,tr}} U^\dag_{\text{tr,tst},B} Z_B,\\
        G_j = U_{\text{tr,tst},B}S_j U_{\text{tr,tst},B}^\dag
        \end{gather}
    where
    \begin{gather}
        S_0 = \mathds{1}  - 2|0\rang\lang0|_{\text{tst, tr,}B},\\ \quad S_j = \mathds{1}  - 2|0\rang\lang0|_{\text{tst}, B} \otimes |\phi_k\rang\lang \phi_k|.
    \end{gather}
    To show the equivalence between $G$ and controlled $G_k$, it suffices to show the equivalence between $\mathcal W S_0 \mathcal W ^\dag$  and $S_k$. Recall that we introduced $\mathcal W_{\text{ind, tr}}$  for the preparation of train states. The action of $\mathcal W$ is given by 
    \begin{equation}
        \mathcal W_{\text{ind, tr}} |i\rang_{\text{ind}}|0\rang_{\text{tr}} = |i\rang_{\text{ind}}|\phi_i\rang_{\text{tr}}.
    \end{equation}
    We have
    \begin{equation}
        \mathds{1}_{\text{in}}\otimes S_0 = \mathds{1}  - 2\sum_{k = 0}^{M-1}|k,0,0,0\rang\lang k, 0,0,0|_{\text{in,tst,tr,}B},
    \end{equation}
    and therefore
    \begin{equation}
    \begin{split}
        \mathcal W S_0 \mathcal W ^\dag &= \mathcal W_{{\text{ind,tr}}}  \left(\mathds{1}  - 2\sum_{k=0}^{M-1}|k,0,0,0\rang\lang k,0,0,0|_{\text{in, tst, tr},B}\right) \mathcal W ^\dag_{\text{ind,tr}} 
        \\&= \mathds{1}  - 2 \sum_{k=0}^{M-1}\mathcal W |k,0,0,0\rang \lang k, 0,0,0|\mathcal W ^\dag \\
         &= \mathds{1}  - 2\sum_k| k\rangle | \phi_k \rangle | 0,0\rang \lang k| \langle \phi_k | \langle 0,0| \\
        &=\sum_k |k\rang \lang k|_\text{in} \otimes \mathds{1}_{\text{tst,tr,}B} - 2\sum_k |k\rang \lang k|_\text{in} \otimes |\phi_k\rang \lang \phi_k| \otimes |0\rang \lang 0|_{\text{tst,}B}\\
    \end{split}
    \end{equation}
    This is
    \begin{equation}
        \mathcal W S_0 \mathcal W  ^\dag =\sum_k |k\rang\lang k| \otimes\Big(
        \mathds{1} - 2|0\rang \lang0|_{\text{tst, }B} \otimes |\phi_k \rang \lang \phi_k|_\text{tr}\Big) 
         = \sum_k|k\rang\lang k|_\text{in} \otimes {S_k}_{\text{tst, tr,}B}  
    \end{equation}
    as required. 

\section{Eigenvectors of $G_j$ for the fidelity based  Q$k$NN} \label{app:G_Eig}
    Recall that $G_j = US_j U^\dag Z_B$ and $U |0 \rang_{\text{tst}} | \phi_{j_{\text{tr}}} \rang |0 \rang_{\text{B}} = |\Psi_j \rang$ 
    This implies,
    \begin{equation}
        U S_j U ^ {\dag} = \mathds{1} - 2 |\Psi_{j}\rang \lang \Psi_{j}|.
    \end{equation}    
   Let 
    \begin{equation}
        |\Psi_{j0}\rang = \frac{1}{2 \alpha_j}  \Big(|\psi\rang_{\text{tr}}|\phi_j\rang_\text{tst} + |\phi_j\rang_{\text{tr}}|\psi\rang_\text{tst}\Big)|0\rang_B \quad \text{and} \quad  
        |\Psi_{j1}\rang = \frac{1}{2 \beta_j}  \Big(|\psi\rang_{\text{tr}}|\phi_j\rang_\text{tst} - |\phi_j\rang_{\text{tr}}|\psi\rang_\text{tst}\Big)|1\rang_B.  
    \end{equation}
    where,
    \begin{equation}
        \alpha_j = \sqrt{\frac12(1 + F_j)} \quad \text{and} \quad 
        \beta_j = \sqrt{\frac12(1 - F_j)}. 
    \end{equation}
    Therefore we have,
    \begin{equation}
        |\Psi_{j}\rang = \alpha_j |\Psi_{j0}\rang + \beta_j |\Psi_{j1}\rang.
    \end{equation}
    Consider the subspace $\mathcal{H} = \text{span}\Big(|\Psi_{j0}\rang, |\Psi_{j1}\rang\Big)$. We have
    \begin{equation}
        Z_B \Big|_{\mathcal{H}} = |\Psi_{j0}\rang \lang \Psi_{j0}|
        - |\Psi_{j1}\rang \lang \Psi_{j1}|, 
    \end{equation}
    and
    \begin{equation}
        U S_j U ^ {\dag} \Big|_{\mathcal{H}} = (1 - 2\alpha_j ^ 2) |\Psi_{j0}\rang \lang \Psi_{j0}|
        + (1 - 2\beta_j ^ 2) |\Psi_{j1}\rang \lang \Psi_{j1}|
        - 2\alpha_k \beta_k(|\Psi_{j1}\rang \lang \Psi_{j0}| + |\Psi_{j0}\rang \lang \Psi_{j1}|).
    \end{equation}
    This implies that
    \begin{equation}     \label{g_j_eq}
        \begin{split}
            G_j \Big|_{\mathcal{H}} &= U S_j U ^ {\dag} Z_B \Big|_{\mathcal{H}} \\
            &= (1 - 2\alpha_j ^ 2) |\Psi_{j0}\rang \lang \Psi_{j0}|
            - (1 - 2\beta_j ^ 2) |\Psi_{j1}\rang \lang \Psi_{j1}|
            - 2\alpha_k \beta_k(|\Psi_{j1}\rang \lang \Psi_{j0}| - |\Psi_{j0}\rang \lang \Psi_{j1}|).
        \end{split}
    \end{equation}

    We can see that this $G_j \Big|_{\mathcal{H}}$ has the same structure as the analogous operator in ~\cite{Mitarai2019}. By using the same proof of correctness of abs-QADC in ~\cite{Mitarai2019} we can see that each $|\Psi_j\rang_{\text{tr,tst}, B}$ can be decomposed into two eigenstates of $G_j$. Let $\alpha_j = \text{sin}(\pi \theta_j)$ for $\theta_j \in [\frac{1}{4}, \frac{1}{2} )$. Substituting this in Eq.~\eqref{g_j_eq}, we get
    \begin{equation}
        G_j \Big|_{\mathcal{H}} = \text{cos}(2 \pi \theta_j) (|\Psi_{j0}\rang \lang \Psi_{j0}| 
        + |\Psi_{j1}\rang \lang \Psi_{j1}|) 
        - \text{sin}(2 \pi \theta_j) (|\Psi_{j1}\rang \lang \Psi_{j0}|
        - |\Psi_{j0}\rang \lang \Psi_{j1}|).
    \end{equation}
    We may write $G_j \Big|_{\mathcal{H}} $ as 
    \begin{equation}
        G_j \Big|_{\mathcal{H}} =
        \begin{bmatrix}
            \text{cos}(2 \pi \theta_j) & \text{sin}(2 \pi \theta_j) \\
            -\text{sin}(2 \pi \theta_j) & \text{cos}(2 \pi \theta_j)
        \end{bmatrix}.
    \end{equation}
    This implies that $G_j$ has eigenvectors 
    \begin{equation}
        |\Psi_{j\pm}\rang = \frac{1}{\sqrt{2}} (|\Psi_{j0}\rang \pm i|\Psi_{j1}\rang),
    \end{equation}
    with eigenvalues $\Psi_{\pm} = e^{\pm i 2\pi \theta_j}$ respectively.
    Now, we can decompose $|\Psi_j \rang$ as
    
    \begin{equation}
        |\Psi_j \rang = \frac{-i}{\sqrt{2}} (e^{i\pi \theta_j} |\Psi_{j+}\rang - e^{-i\pi\theta_j} |\Psi_{j-}\rang),
    \end{equation}
    as required. 
    
    
    
\section{Action of $H$ as controlled $H_j$ for the dot product based  Q$k$NN} \label{actH}
    Recall that
    \begin{equation}\begin{split}
        H &= V_{\text{in, data},B} \ \mathcal S_{0_{\text{data,B}}} \ \mathcal  V^\dag_{\text{in, data},B} \ Z_B,\\
        H_j &= (\mathds{1} - 2 |\Psi_j \rang \lang \Psi_j |_{\text{data,B}}) Z_B 
    \end{split}\end{equation}

    Then,
    \begin{equation}
        \begin{split}
            H &= V_{\text{in,data,B}}
            \Big(\mathds{1}_{\text{in,data,B}} - \mathds{1}_{\text{in}} \otimes 2 |0\rang \lang 0|_{\text{data,B}}\Big) V_{\text{in,data,B}} ^ {\dag} \Big(\mathds{1}_{\text{in,data}} \otimes Z_B\Big) \\
            &= \Big(\mathds{1}_{\text{in,data,B}}
            - \sum \limits_{k = 0} ^ {M - 1} 2 |k\rang \lang k|_{\text{in}} \otimes  |\Psi\rang  \lang \Psi |_{\text{data,B}} \Big)  \Big(\mathds{1}_{\text{in,data}} \otimes Z_B\Big) \\
            & =\sum \limits_{k = 0} ^ {M - 1} |k \rang \lang k |_{\text{in}} \otimes \Big( \mathds{1}_{\text{data},B} - 2 |\Psi_j \rang \lang \Psi_j |_{\text{data},B} \Big) \Big(\mathds{1}_{\text{in,data}} \otimes Z_B\Big) \\
            & =\sum \limits_{k = 0} ^ {M - 1} |k \rang \lang k |_{\text{in}} \otimes {H_k}_{\text{data,B}}.
        \end{split}
    \end{equation}    
 
\section{Eigenvectors of $H_j$ for the dot product based  Q$k$NN} \label{eigH}
    Recall that $H_j = \Big( \mathds{1} - 2 |\Psi_{j}\rang \lang \Psi_{j}| \Big) Z_B$. Let 
    \begin{equation}
        |\Psi_{j0}\rang = \frac{1}{2 \alpha_j} \Big( |v\rang_{\text{data}} + |u_j\rang_{\text{data}}\Big) |0\rang_B \quad \text{and} \quad 
        |\Psi_{j1}\rang = \frac{1}{2 \beta_j} \Big( |v\rang_{\text{data}} - |u_j\rang_{\text{data}}\Big) |1\rang_B,
    \end{equation}
    where,
    \begin{equation}
        \alpha_j = \sqrt{\frac12(1 + X_j)} \quad \text{and} \quad \beta_j = \sqrt{\frac12(1 - X_j)}. 
    \end{equation}
    So we may write,
    \begin{equation}
        |\Psi_{j}\rang = \alpha_j |\Psi_{j0}\rang + \beta_j |\Psi_{j1}\rang.
    \end{equation}
    Consider the subspace $\mathcal{H} = \text{span}\Big(|\Psi_{j0}\rang, |\Psi_{j1}\rang\Big)$. Then
    \begin{equation}
        Z_B \Big|_{\mathcal{H}} = |\Psi_{j0}\rang \lang \Psi_{j0}|
        - |\Psi_{j1}\rang \lang \Psi_{j1}|. 
    \end{equation}
    We have
    \begin{equation}
        \Big( \mathds{1} - 2 |\Psi_{j}\rang \lang \Psi_{j}| \Big) \Big|_{\mathcal{H}} = (1 - 2\alpha_j ^ 2) |\Psi_{j0}\rang \lang \Psi_{j0}|+ (1 - 2\beta_j ^ 2) |\Psi_{j1}\rang \lang \Psi_{j1}|- 2\alpha_k \beta_k(|\Psi_{j1}\rang \lang \Psi_{j0}| + |\Psi_{j0}\rang \lang \Psi_{j1}|).
    \end{equation}
    This implies
    \begin{equation}    \label{h_j_eq}
        \begin{split}
            H_j \Big|_{\mathcal{H}} = (1 - 2\alpha_j ^ 2) |\Psi_{j0}\rang \lang \Psi_{j0}|- (1 - 2\beta_j ^ 2) |\Psi_{j1}\rang \lang \Psi_{j1}| - 2\alpha_k \beta_k(|\Psi_{j1}\rang \lang \Psi_{j0}| - |\Psi_{j0}\rang \lang \Psi_{j1}|).
        \end{split}
    \end{equation} 
    The result immediately follows by comparing Eq.~\eqref{h_j_eq} with Eq.~\eqref{g_j_eq}.

    \begin{figure}
            \begin{center}
                \includegraphics[page=9]{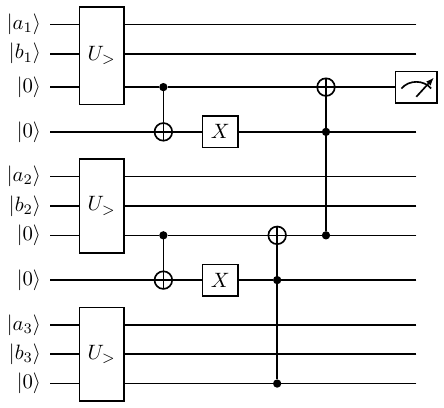}
                \caption{Circuit for implementing $J$ gate for $3$ qubit bit strings. Note that by replacing the $U_{>}$ with $U_{\neq}$ gates, we can implement $D ^ {(i)}$ gates.}
                \label{umain}
            \end{center}
    \end{figure}

    \begin{figure*}[t!]
    \centering
    \begin{subfigure}[b]{0.5\textwidth}
        \centering
    \includegraphics[page=8]{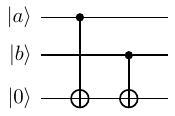}
                \caption{Circuit for $U_{\neq}$}
                \label{U_neq}
    \end{subfigure}%
    ~ 
    \begin{subfigure}[b]{0.5\textwidth}
        \centering
        \includegraphics[page=7]{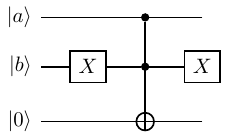}
            \caption{Circuit for $U_>$.}
            \label{fig:U_geq}
    \end{subfigure}
    \caption{Circuits used in bitstring comparison.}
\end{figure*}

\section{Bit-string comparison} \label{app:bs_comp}
    In this section, we give a basic example of a circuit capable of comparing two-qubit bit string states. The $J$ gate, mentioned in Eq.~\eqref{jgate}, for $1$ qubit bit strings, can be implemented by the $U_{>}$ gate given in Fig. \ref{fig:U_geq}.
    
    Using $U_>$ as a subroutine, in Fig. \ref{umain} we give an example of a circuit capable of comparing $3$-qubit bit strings $|a\rangle = |a_1\rangle |a_3\rangle |a_3\rangle$ and $|b\rangle = |b_1\rangle |b_2\rangle |b_3\rangle$.
    
    This circuit compares the bits from left to right using the $U_{>}$ gate, that is, first $a_3$ and $b_3$ are compared, then $a_2$ and $b_2$ and so on. At any point, if $U_{>}$ returns $1$, then the final output will be $1$.
    
    For executing the $D ^ {(i)}$ used in Eq.~\eqref{dgate}, we replace the $U_{>}$ gate with the $U_{\neq}$ gate given in Fig. \ref{U_neq}. This gate returns $1$ if the input bits are unequal. By using $U_{\neq}$ in the circuit $U$, we can check, from left to right, if each pair of qubits are unequal or not. At any point, if we see an unequal pair of qubits, the circuit outputs $1$, else 0. Different versions of these circuits and extension to $n$ qubits are explained in \cite{Oliveira2006}.

\newpage

\end{document}